\documentclass{pasa}

\usepackage[dvipsnames]{xcolor}
\usepackage{upgreek}
\newcommand\ion[2]{#1$\;${\scshape{#2}}}
\newcommand\micron{\rm{\upmu m}}

\title[SPICA: The Rise of Metals and Dust]{\textit{SPICA} and the Chemical Evolution of Galaxies:\\ The Rise of Metals and Dust}
\author[Fern\'andez-Ontiveros et al.]{J.A.~Fern\'andez-Ontiveros$^{1,2,3\thanks{Email: {\sf j.a.fernandez.ontiveros@gmail.com, jafo@iac.es}}}$, L.~Armus$^4$, M.~Baes$^5$, J.~Bernard-Salas$^6$, A.D.~Bolatto$^7$, J.~Braine$^8$, L.~Ciesla$^9$, I.~De~Looze$^{10}$, E.~Egami$^{11}$, J.~Fischer$^{12}$, M.~Giard$^{13,14}$, E.~Gonz\'alez-Alfonso$^{15}$, G.L.~Granato$^{16}$, C.~Gruppioni$^{17}$, M.~Imanishi$^{18}$, D.~Ishihara$^{19}$, H.~Kaneda$^{19}$, S.~Madden$^{9}$, M.~Malkan$^{20}$, H.~Matsuhara$^{21}$, M.~Matsuura$^{22}$, T.~Nagao$^{23}$, F.~Najarro$^{24}$, T.~Nakagawa$^{21}$, T.~Onaka$^{25}$, S.~Oyabu$^{19}$, M.~Pereira-Santaella$^{26}$, I.~P\'erez~Fournon$^{1,2}$, P.~Roelfsema$^{27,28}$, P.~Santini$^{29}$, L.~Silva$^{16}$, J.-D.T.~Smith$^{30}$, L.~Spinoglio$^3$, F.~van~der~Tak$^{27,28}$, T.~Wada$^{21}$, \and R.~Wu$^{31}$}

\jid{PASA}
\doi{10.1017/pas.\the\year.xxx}
\jyear{\the\year}

\usepackage[authoryear]{natbib}
\bibpunct{(}{)}{;}{a}{}{,}
\usepackage{aas_macros}
\usepackage{hyperref}
\hypersetup{colorlinks,citecolor=blue,linkcolor=blue,urlcolor=blue}

\usepackage{newtxtext}
\usepackage[libertine]{newtxmath}

\usepackage[T1]{fontenc}

\begin{document}%

\begin{abstract}
The physical processes driving the chemical evolution of galaxies in the last $\sim 11\, \rm{Gyr}$ cannot be understood without directly probing the dust-obscured phase of star-forming galaxies and active galactic nuclei. This phase, hidden to optical tracers, represents the bulk of star formation and black hole accretion activity in galaxies at $1 < z < 3$. Spectroscopic observations with a cryogenic infrared (IR) observatory like \textit{SPICA} will be sensitive enough to peer through the dust-obscured regions of galaxies and access the rest-frame mid- to far-IR range in galaxies at high-$z$. This wavelength range contains a unique suite of spectral lines and dust features that serve as proxies for the abundances of heavy elements and the dust composition, providing tracers with a feeble response to both extinction and temperature. In this work, we investigate how \textit{SPICA} observations could be exploited to understand key aspects in the chemical evolution of galaxies: the assembly of nearby galaxies based on the spatial distribution of heavy element abundances, the global content of metals in galaxies reaching the knee of the luminosity function up to $z \sim 3$, and the dust composition of galaxies at high-$z$. Possible synergies with facilities available in the late 2020s are also discussed.
\end{abstract}

\maketitle%

{\bf Preface}

\vspace{0.5cm}
\noindent
The following set of papers describe in detail the science goals of the future \textit{Space Infrared telescope for Cosmology and Astrophysics} (\textit{SPICA}). The \textit{SPICA} satellite will employ a 2.5-m telescope, actively cooled to around $6\,\rm{K}$, and a suite of mid- to far-IR spectrometers and photometric cameras, equipped with state of the art detectors. In particular the \textit{SPICA} Far-Infrared Instrument (SAFARI) will be a grating spectrograph with low ($R = 300$) and medium ($R \simeq 3\,000$--$11\,000$) resolution observing modes instantaneously covering the $35$--$230\, \rm{\micron}$ wavelength range. The \textit{SPICA} Mid-Infrared Instrument (SMI) will have three operating modes: a large field of view ($12' \times 10'$) low-resolution $17$--$36\, \rm{\micron}$ spectroscopic ($R \sim 50$--$120$) and photometric camera at $34\, \rm{\micron}$, a medium resolution ($R \simeq 2\,000$) grating spectrometer covering wavelengths of $17$--$36\, \rm{\micron}$, and a high-resolution echelle module ($R \simeq 28\,000$) for the $12$--$18\, \rm{\micron}$ domain. A large field of view ($80'' \times 80''$), three channel ($110$, $220$ and $350\, \rm{\micron}$), polarimetric camera will also be part of the instrument complement. These articles will focus on some of the major scientific questions that the \textit{SPICA} mission aims to address, more details about the mission and instruments can be found in \citet{roe17}.


\section{INTRODUCTION}
\label{sec:intro}

Heavy elements (i.e. elements heavier than He, also known as metals) carry the signature of galaxy evolution. The study of their content and composition in galaxies allows us to look back at the history of the internal recycling of matter: gas accretion feeding star formation episodes and nuclear activity produced by supermassive black holes (SMBH), the enrichment of the gas in the interstellar medium (ISM) or the depletion of heavy elements into dust. The ensemble of the processes involved, known as the ``baryon cycle'' \citep{dav12}, are invoked to explain fundamental characteristics observed in the present day Universe 
such as the mass metallicity relation, where the highest metallicities are found in the most massive galaxies \citep{tre04,gal06}.

In this scenario, the interstellar and intergalactic media --\,ISM and IGM, respectively\,-- are progressively enriched with heavy elements after a series of matter cycling. These are mixed in the gas-phase of the ISM and IGM, but are also the main components of the solid-phase in the form of dust grains \citep{dra03}. Despite their small contribution in terms of mass fraction ($\lesssim 1\%$ of the galaxy mass), heavy elements and dust are fundamental components with a strong influence on the observational properties of galaxies and a main role in the physical and chemical processes that take place in the ISM, e.g. absorption of the optical and UV radiation and re-emission into the infrared (IR), ISM cooling favouring the cloud collapse and subsequent star formation.

Therefore, to understand the physical mechanisms driving the evolution of galaxies, it is essential to probe the products of the baryon cycle over cosmic time. Since the vast majority of star formation and SMBH accretion activity in the universe occurs during a dust embedded phase in the evolution of galaxies \citep[see][]{mad14}, the content and composition of metal and dust cannot be accurately derived from tracers in the rest-frame obscured optical range. The use of tracers able to peer into highly obscured environments is mandatory to investigate the content of metals and dust from the peak of star formation activity and SMBH accretion at $z \sim 1$--$3$ to the Local Universe \citep[e.g.][]{boy98,fra99,mad14,man16}. Thus, the answers to fundamental questions in the field of galaxy evolution rely on the contribution of an IR observatory: how are heavy elements and dust produced? how is dust destroyed? Can we describe quantitatively the cycle of matter in galaxies and the physical processes involved? What fundamental physical processes determine the chemical evolution of galaxies?

The present work is focused on the role that a cryogenic IR observatory like the \textit{SPace Infrared telescope for Cosmology and Astrophysics}\footnote{\url{http://www.spica-mission.org}} (\textit{SPICA}; \citealt{swi09}, \citealt{nak14}, \citealt{sib15}, \citealt{roe17}), with its $2.5\, \rm{m}$ primary mirror actively cooled to around $6\, \rm{K}$, would play to understand the chemical evolution of galaxies from $z \sim 3$ till the present time. The \textit{SPICA} observatory has been proposed by a European-Japanese collaboration as the fifth medium-sized class (M5) mission of the European Space Agency (ESA) Cosmic Vision program. In this paper, enclosed in the \textit{SPICA} white paper series on galaxy evolution, we explore the power of \textit{SPICA} in detecting lines and features which can allow robust measurements of the metallicity and dust composition in galaxies. We refer to \citet{spi17} for the general overview on \textit{SPICA} studies of galaxy evolution and for the description of the power of IR line diagnostics.
The article is organized as follows: in Section\,\ref{chevol}, we address the key scientific questions that we want to investigate; the main observables, abundance tracers, and dust features available in the mid- to far-IR range are described in Section\,\ref{observables}; in Section\,\ref{buildup} we discuss the role that \textit{SPICA} would play to address the above science questions; 
the possible synergies with future facilities available in the late 2020s are described in Section\,\ref{syner}; finally, the main conclusions of this study are summarised in Section\,\ref{sum}.

\section{THE CHEMICAL EVOLUTION OF GALAXIES}\label{chevol}

The total content of dust and metals in galaxies is determined by the balance among star formation, the different episodes of gas accretion and ejection (produced by gravity torques and gas inflow motions, and winds and outflows, respectively), and the role of active galactic nuclei (AGN). The influence of these physical processes in the ISM chemistry and its evolution over cosmic time --\,from the build-up of metals and dust in galaxies till the present time\,-- are still a matter of discussion. However, the past history of galaxy evolution is recorded in the global content and distribution of metals and dust. These are the main products of the matter cycle and show their imprint in the IR spectrum, in the form of continuum emission, spectral lines, and dust spectral features. Moreover, their spatial distribution across the galaxy may also reflect the past activity of the baryon cycle. Thus, a consistent picture describing the role of the different physical mechanisms in the evolution of galaxies should explain, at once, how the global content of metals and dust in galaxies evolves with cosmic time, and how galaxies were assembled.

In this context, we identify three main questions in the field of chemical evolution whose answers are key to understand how galaxies have evolved in the last $\sim 11\, \rm{Gyr}$: \textit{i)} how are metals distributed within galaxies? \textit{ii)} how did the global content of metals in galaxies evolve with cosmic time? and \textit{iii)} how did metals locked into dust grains evolve from high-$z$ till the present time?

\subsection{The Assembly of Galaxies}\label{nearby}
Current models of galaxy formation predict that galaxy discs form through accretion of matter in an inside-out growth, with a combination of inflow and outflow events regulating the process of metal enrichment \citep{fu09,dav12,ang14}. The accreted gas reaches higher densities in the interior of the galaxy, leading to higher star formation rates and a faster reprocessing of gas, which results in higher metallicities in the bulge and in the inner part of discs. This is followed by a decreasing metallicity towards outer regions, populated by younger stars formed from poorly enriched material \citep[e.g.][]{gd15}.
While outflow activity is potentially capable of transferring the high abundance of metals to the halo or even the IGM \citep{opp06}, and subsequent inflows can further recycle them, observations have to be performed to ascertain the history of chemical evolution and the way in which outflows and inflows have shaped it (see also \mbox{\citealt{gon17}}).
Since the spatial distribution contain the imprint of gas inflow and outflow episodes throughout the lifetime of these galaxies, tracing the spatial gradients of heavy element abundances in present day galaxies provides a unique insight into their evolutionary history.

\subsection{The Evolution of Metallicity in Galaxies}\label{fmr}
\begin{figure}
  \begin{center}
  \includegraphics[width=\columnwidth]{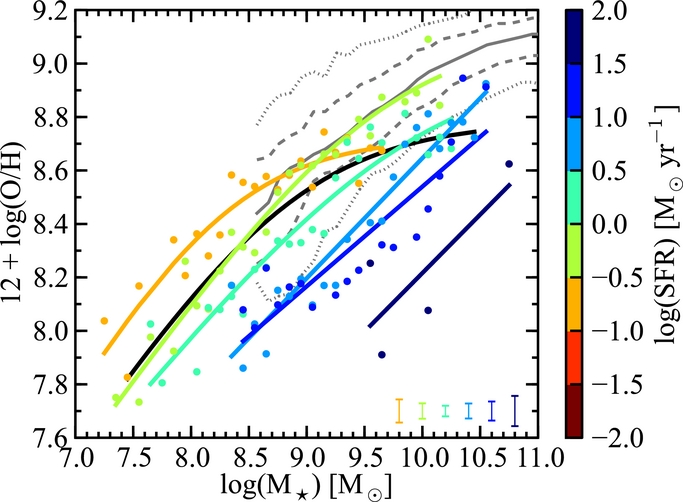}
  \caption{
The mass-metallicity relation determined from direct abundance measurements based on optical fine-structure lines for $\sim 200\,000$ star-forming galaxies selected from the Sloan Digital Sky Survey ($0.027 < z < 0.25$), grouped in different bins of SFR. The colour coded circles correspond to stacks of galaxies for each bin in SFR (mean error is indicated in the lower right). The coloured solid lines show asymptotic logarithmic fits to this relation for each bin in SFR, and the black-solid line in the centre of the figure shows the global fit. These are compared with the mass-metallicity relation from \citet{tre04}, based on the strong-line method (the solid, dashed, and dotted grey lines correspond to the median, 68\%, and 95\% contour of the relation, respectively). Figure adapted from \citet{and13}.}\label{metalsfr}
\end{center}
\end{figure}

The ISM of massive galaxies shows a higher content of heavy elements when compared to less massive ones, as was discovered early by \citet{leq79}. Later \citet{man10} and \citet{lar10} incorporated and additional dependency on the SFR required to describe the evolution of galaxies up to $z \sim 3.7$. Thus, the impact of the baryon cycle determines the relation among fundamental properties of galaxies such as the gas metallicity, the stellar mass, and the SFR. This is illustrated in Figure\,\ref{metalsfr}, where the mass-metallicity relation is obtained for galaxies in different bins of SFR \citep{and13}. The metallicities in Fig.\,\ref{metalsfr} are based on the direct method, i.e. using the weak auroral lines to determine the gas temperature, the latter is one of the main uncertainties of the strong-line method based on the calibration of bright emission lines sensitive to the metallicity \citep[e.g.][see Section\,\ref{observables} for further details]{tre04}. Thereafter \citet{bot13} interpreted the dependence on the SFR as a by-product of the molecular gas content, via the Schmidt-Kennicutt relation \citep{sch59,ken98}. In this scenario, the chemical enrichment is driven by the increase of the specific SFR due to the larger disc gas fractions in galaxies at high-$z$ \citep{bot16a,bot16b}, and can be triggered by the accretion of pristine gas from the cosmic web and/or mergers with metal-poor companions \citep{hun16a,hun16b}. On the other hand, outflows of enriched material dominate at low-redshift, explaining the correlation with the stellar mass \citep[e.g.][]{pee11,dav11,som15}.

In the present-day picture, the relation between galaxy mass and gas metallicity is a by-product of the internal cycle of matter in galaxies during their evolution with cosmic time \citep{dav12}, and it is due to several reasons: \textit{i)} starburst wind outflows in low-mass galaxies eject the metal-enriched material to the halo and the IGM due to the shallow gravitational potential well, resulting in a less efficient enrichment compared to massive galaxies \citep{tre04}; \textit{ii)} the ``galaxy downsizing'' scenario, i.e. low-metallicity galaxies in the Local Universe are associated with a high specific SFR and are found in low-mass systems, which are still in an early evolutionary stage because they have converted only a small fraction of their gas mass into stars \citep[e.g.][]{bro07}; \textit{iii)} dependence of the shape of the initial mass function (IMF) with the galaxy mass \citep{koe07}; and \emph{iv)} infall of metal-poor gas from the halo and/or the IGM \citep{fin08,bro09}.

If the same physical processes are in place in the Local Universe and at high-$z$, then the chemical evolution of galaxies should be governed by a reduced number of fundamental properties. Diverse models of galaxy formation and evolution predict manifold evolutionary patterns of the mass-metallicity relation as a function of redshift. These patterns lead to different properties in the modelled galaxies such as the slope of the mass-metallicity relation and its scaling with redshift, scatter around the mass-metallicity relation due to fluctuations in the baryon cycle activity, and the mass fraction of atomic and molecular gas \citep[e.g.][]{dav17}. So far, most of the models reproduce the growth of the stellar mass in galaxies over cosmic time, but the global content of heavy elements and the gas fraction are largely unconstrained. Therefore, observational data of galaxies from high-$z$ to the Local Universe are required to test these properties, discriminate among the different cosmological simulations, and understand the physical processes driving the chemical evolution in galaxies.

\subsection{The Build-Up of Cosmic Dust}\label{cosmicdust}
Heavy elements produced in the baryon cycle are partially locked into dust grains, and constitute the solid-phase of the ISM \citep{dra03,pei10}. The dust has a strong influence on the observational properties of galaxies, with a main role in the heating and cooling processes \citep{spi48,bor91,hoo01}, and as catalyst of chemical processes in the ISM (e.g. the formation of H$_2$ driving the molecular chemistry). Dust grains absorb the UV light and re-radiate it in the IR, shielding the dense gas and by these means triggering the formation of molecular clouds where new stars are born. The grain surfaces and the polycyclic aromatic hydrocarbons (PAHs) participate in a large number of chemical reaction networks in different phases of the ISM \citep{bak98}, and also provide photoelectric heating of gas in photo-dissociation regions \citep{dra11,bak94}. Moreover, dust cooling is able to alter the shape of the IMF by favouring the cloud fragmentation, thus inhibiting the formation of massive stars and fostering the creation of low-mass stars \citep{omu05}. On the other hand, dust grains suffer from processing and destruction due to intense radiation fields, sputtering (impacts with gas ions), shattering (grain-grain collisions), shocks due to SNe explosions, astration into newly formed stars, and photo-processing by cosmic rays and UV/X-ray fields in the vicinity of AGN.

Little is known on the dust chemistry and its evolution over cosmic time. Even in the Milky Way and its neighbourhood, fundamental aspects in the astrochemistry of dust are not well understood: the dust composition in star forming regions \citep{pei10}, the ISM \citep[e.g.][]{nie12}, and planetary nebulae (PNe; \citealt{cle89,ber09a}); the depletion of elements onto dust grains \citep[e.g.][]{del09,sim11}; the dust-to-gas ratios, grain size distribution, and dust destruction \citep{mal88,sta99,hoo13,asa13b}; the physical mechanisms involved in the dust-phase catalysis within the ISM and subsequent enrichment of the gas-phase \citep{dul13}. Nevertheless, the dust composition and its properties are critical ingredients to describe the evolution of galaxies \citep{rem14}, while the uncertainties mentioned above escalate significantly for studies of galaxies at high-$z$ \citep[e.g.][]{pop17}. Currently, models describing the chemical evolution in galaxies \citep[e.g.][]{asa13a} rely on dust studies in local metal-poor dwarfs \citep[e.g.][]{rem15}, which are usually adopted as a benchmark case for the ISM in galaxies at high-$z$. However, there is firm observational evidence suggesting that the dust properties are not universal, e.g. chemical composition, gas-to-dust and dust-to-metal ratios, and grain size distribution \citep[see][and references therein]{zhu16}. The characteristics of dust in the early Universe may be quite different from those seen in local metal-poor dwarfs, thus probing the dust properties in galaxies at high-$z$ is the only way to constrain the solid phase of the ISM in chemical evolution models.

\section{METALLICITY TRACERS IN THE INFRARED}\label{observables}

An accurate description of the chemical evolution in galaxies requires reliable tracers to quantify the abundances of the different chemical species in the ISM. Specifically, it is essential to use appropriate tracers able to peer into the dust-embedded phase in the evolution of galaxies, which represents the bulk of the star formation and black hole accretion activity \citep{mad14,man16}.

So far, the vast majority of studies probing the abundances of heavy elements and their evolution with redshift rely on the determination of oxygen abundances using optical line diagnostics. Among them one can distinguish two main types of diagnostics, those based on the determination of the electron temperature using the weak auroral lines (the direct method; e.g. \citealt{izo94,izo97,izo99,ken03,hag08}), and those based on the calibration of bright emission-line ratios sensitive to the gas metallicity \citep{pag79,mcg91,pil01a,pil05,dor05}. The main limitations that affect optical-based tracers are: \textit{i)} dust obscuration, which restricts the abundance measurements to the unobscured regions in galaxies; \textit{ii)} poor determination of the electron temperature, which is based on temperature-sensitive auroral lines (e.g. [\ion{O}{iii}]$\lambda 4363$, [\ion{N}{ii}]$\lambda 5755$), typically faint and difficult to detect \citep{bre05,mou10}; \textit{iii)} systematic discrepancies in the calibration of strong-line abundance diagnostics, based on the brightest collisionally-excited lines such as [\ion{O}{iii}]$\lambda 4959, 5007$, [\ion{N}{ii}]$\lambda 6548, 6584$, [\ion{S}{ii}]$\lambda 6716, 6731$, H$\alpha$, and H$\beta$. Although these calibrations provide reliable abundances in relative terms, they show large systematic discrepancies (up to $\sim 0.7\, \rm{dex}$) in the absolute oxygen abundances derived among the various empirical and theoretical calibrations for the line ratios \citep[see][and references therein]{kew08}. This is especially the case at high-$z$ \citep{kew06}, where physical conditions in the ISM can be significantly different than those in nearby galaxies. Furthermore, they also show a different turnover in the mass-metallicity relation when compared to direct metallicity determinations (see black- and grey-solid lines in Fig.\,\ref{metalsfr}; \citealt{and13}); \textit{iv)} temperature fluctuations across the nebula --\,e.g. produced by variations in the chemical abundances or densities\,-- can lead to an error in the abundance estimates based on optical lines of a factor $\sim 4$ \citep{dor13}. Moreover, except for the case of oxygen which shows strong emission lines for its main chemical species in the optical (e.g. [\ion{O}{i}]$\lambda 6300, 6364$, [\ion{O}{ii}]$\lambda 3726, 3729$, [\ion{O}{iii}]$\lambda 4959, 5007$), the rest of element abundances have to rely on uncertain ionisation correction factors to account for the unobserved stages of ionisation. Commonly adopted metallicity calibrations (i.e. indirect methods), based on the ratios of strong emission lines in the optical and UV range, are: the index R$_{23} \equiv$ ([\ion{O}{ii}]$\lambda 3727$ + [\ion{O}{iii}]$\lambda 4959, 5007$)/H$\beta$ \citep{pil01b,pil05}; the index N$_2 \equiv$ [\ion{N}{ii}]$\lambda 6584$/H$\alpha$ \citep{pp04}; the index O$_3$N$_2 \equiv$ ([\ion{O}{iii}]$\lambda 5007$/H$\beta$)/([\ion{N}{ii}]$\lambda 6584$/H$\alpha$) \mbox{\citep{pp04}}; a combination of [\ion{N}{ii}]$\lambda 6584$, H$\alpha$, and [\ion{S}{ii}]$\lambda 6717, 6731$ \citep{dop16}; in the UV, the index C$_{43}\,= \log(($\ion{C}{iv}$\,\lambda 1549$ + \ion{C}{iii}]$\,\lambda 1909)/$\ion{He}{ii}$\,\lambda 1640)$ for AGN \citep{dor14}; and a combination of the indices C$_3$O$_3\, = \log(($\ion{C}{iv}$\,\lambda 1549$ + \ion{C}{iii}]$\,\lambda 1909)/$[\ion{O}{iii}]$\,\lambda 1664)$, C$_{34}\,= \log(($\ion{C}{iv}$\,\lambda 1549$ + \ion{C}{iii}]$\,\lambda 1909)/$\ion{H}{i}$)$ (where \ion{H}{i} corresponds to a hydrogen recombination line, e.g. H$\beta$), and C$_3$C$_4\, = \log($\ion{C}{iii}]$\,\lambda 1909/$\ion{C}{iv}$\,\lambda 1549)$ for star forming regions \citep{p-a17}.

\subsection{Why IR tracers?}\label{whyIR}
The use of robust metallicity tracers almost independent of the dust extinction, radiation field, and of the gas density are crucial to study the chemical evolution of galaxies. For instance, the metallicity inferred from the dust content of high-$z$ sub-millimetre galaxies (SMG), determined with \textit{Herschel} far-IR photometry, is more than an order of magnitude higher when compared to gas metallicity measurements based on optical nebular lines \citep{san10}. This discrepancy is most probably caused by dust obscuration. Since SMG are compact and dust-rich, most of the ISM is optically thick at visual wavelengths. Thus, the optical nebular lines only probe the outer parts of the star-forming regions, which are probably hardly enriched with heavy elements.
\begin{figure*}
  \begin{center}
  \includegraphics[width=0.5\textwidth]{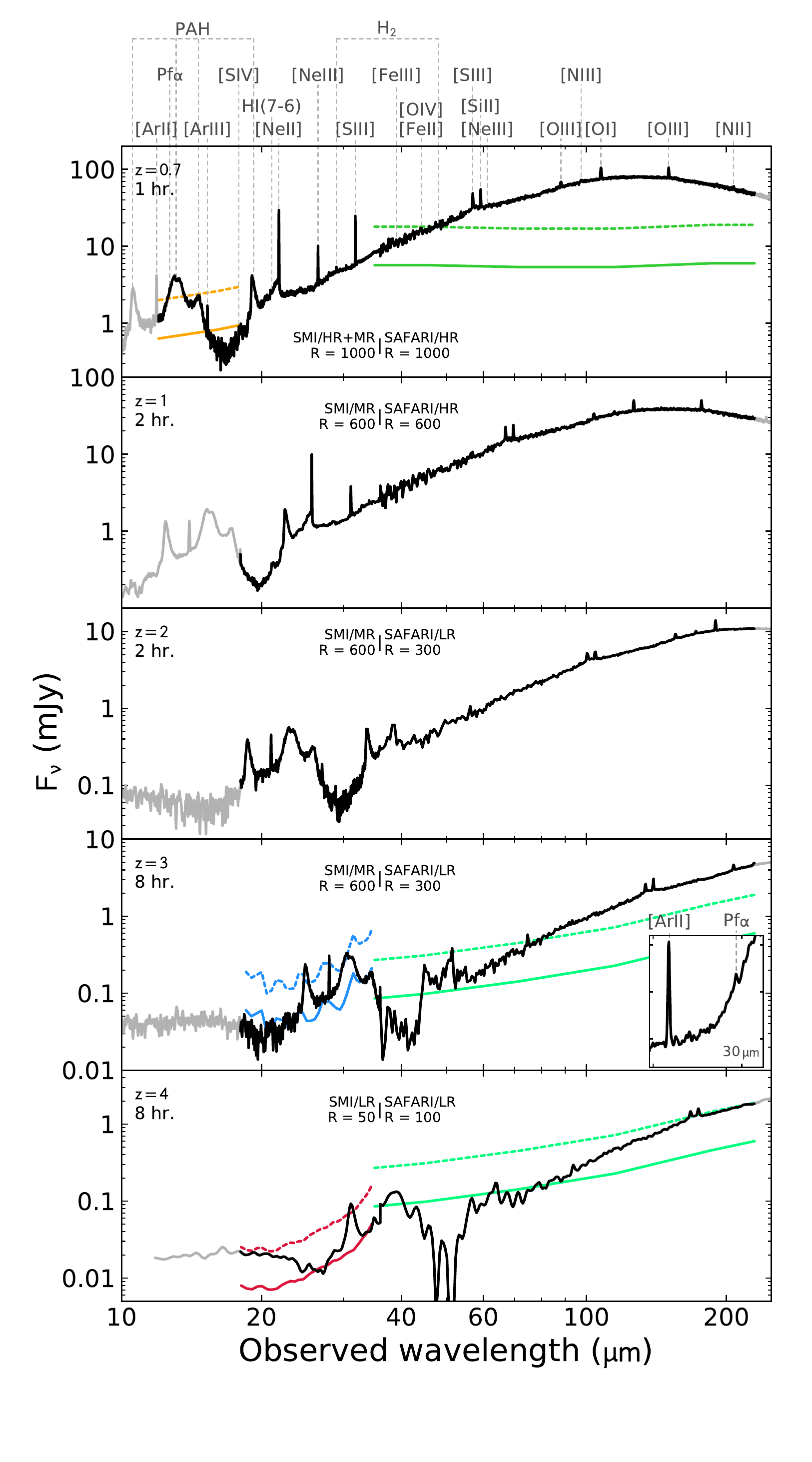}~
  \includegraphics[width=0.5\textwidth]{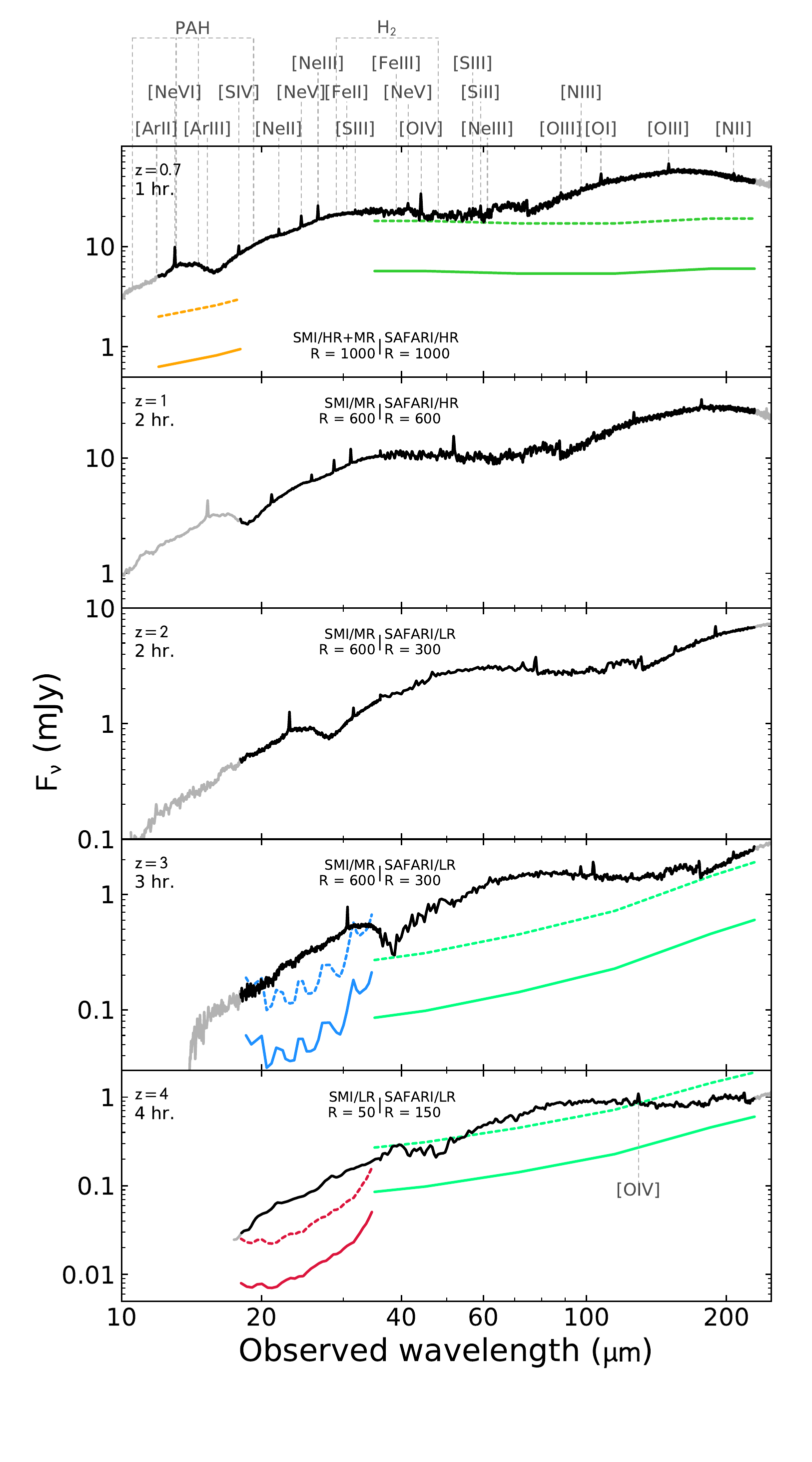}
  \caption{Simulated spectra for a starburst galaxy (left) and an AGN (right) with a luminosity of $2 \times 10^{12}\, \rm{L_\odot}$ in the $z = 1$--$4$ range, assuming the spectral characteristics of M82 and NGC\,1068, respectively, observed with \textit{ISO}/SWS + LWS and \textit{Spitzer}/IRS. The spectra are simulated at $z = 1, 2, 3$ and $4$, from top to bottom. In each case, white noise was added to the spectra with an amplitude of $1 \sigma / \sqrt{t}$, $t$ being the exposure time assumed (indicated in the upper-left corner of each panel). Finally, the SMI and SAFARI spectral ranges were binned to the spectral resolution values indicated in each frame. The \textit{SPICA} $5 \sigma$ sensitivities for SMI (MR mode in blue, $R \sim 1500$; LR mode in red, $R \sim 100$) and SAFARI (in green, HR mode in the upper panels at $z = 0.7$; LR mode, $R \sim 300$, in the lower panels) are indicated for integration times of 1 hour (dotted lines) and 10 hours (solid lines). The Pfund-$\alpha$ would be detected in the $1.5 < z \lesssim 3$ range (see inset on the left panel at $z = 3$).}\label{simul}
\end{center}
\end{figure*}

Abundance determinations based on mid- and far-IR lines offer two great advantages when compared to those derived from UV and optical lines: \textit{i)} IR lines are almost free from dust extinction; and \textit{ii)} they have a small dependency on the electron temperature \citep[e.g.][]{ber01}. The latter could be the origin of a known discrepancy between optical and IR tracers, where IR-based abundances tend to be higher when compared to those obtained using optical tracers \citep[see][]{ver02,dor13}. A comprehensive explanation of this discordance is still lacking, and therefore there is not a consensus in the field on what tracer (optical or IR) is more reliable. However, a possible explanation is given in \citet{dor13}: temperature variations within the nebula that not resolved spatially in the spectra --\,e.g. produced by spatial variations in the chemical abundances\,-- can lead to a systematic underestimation of the abundances derived from optical lines using the direct method by a factor $\lesssim 4$ ($0.6\, \rm{dex}$). For instance, this is the case of integrated spectra in galaxies. If confirmed, this result implies that tracers based on IR lines can be much more accurate, since their dependency on the electron temperature is feeble. Unfortunately, there are few galaxies where both methods can be applied and compared. Sensitive IR observations with the \textit{SPICA} observatory for a sample of galaxies with available optical metallicity estimates would be essential to shed light on this issue and develop new and reliable abundance tracers.

In the Local Universe, \textit{SPICA} would have access to several ionic fine-structure lines in the rest-frame mid- and far-IR spectra of galaxies. Thus, abundance diagnostics for heavy elements will be mainly based on ratios of these lines, which can be calibrated using photoionisation models such as \textsc{Cloudy} \citep{fer17} and \textsc{Mappings}\footnote{\url{https://mappings.anu.edu.au/code}}. The weak dependence with density can be easily constrained using density-sensitive ratios such as [\ion{N}{ii}]$_{122 \rm{\micron}}$ to [\ion{N}{ii}]$_{205 \rm{\micron}}$, [\ion{O}{iii}]$_{52 \rm{\micron}}$ to [\ion{O}{iii}]$_{88 \rm{\micron}}$, [\ion{S}{iii}]$_{18.7 \rm{\micron}}$ to [\ion{S}{iii}]$_{33.5 \rm{\micron}}$, [\ion{Ne}{iii}]$_{15.6 \rm{\micron}}$ to [\ion{Ne}{iii}]$_{36.0 \rm{\micron}}$, and [\ion{Ne}{v}]$_{14.3 \rm{\micron}}$ to [\ion{Ne}{v}]$_{24.3 \rm{\micron}}$ \citep{cro13}. Furthermore, the hardness of the radiation field can also be constrained using the line ratios of [\ion{N}{iii}]$_{57 \rm{\micron}}$ to [\ion{N}{ii}]$_{122, 205 \rm{\micron}}$, [\ion{Ne}{iii}]$_{15.6 \rm{\micron}}$ to [\ion{Ne}{ii}]$_{12.8 \rm{\micron}}$, and [\ion{O}{iv}]$_{25.9 \rm{\micron}}$ to [\ion{O}{iii}]$_{52, 88 \rm{\micron}}$. On the other hand, the effects on the fine-structure lines produced by the combination of high far-infrared optical depths and high ionisation parameters found in luminous and ultraluminous IR galaxies (LIRGs and ULIRGs, respectively; \citealt{gra11}, \citealt{fis14}, \citealt{gon15}) can also be calibrated with photoionisation models (e.g. \mbox{\citealt{abe09}}, \citealt{gra11}, \citealt{fis14}).

Besides the gas-phase abundances, a sensitive telescope in the mid- to far-IR range would be able to probe the composition of dust in galaxies at high-$z$ through several solid-state features in the spectrum, such as forsterite (Mg$_2$SiO$_4$; $23$, $33$, $69\, \rm{\micron}$), water ice at $44$ and $62\, \rm{\micron}$, MgS and FeS at $\sim 30\, \rm{\micron}$, FeO and SiO$_2$ at around $21\, \rm{\micron}$, iron-bearing crystalline silicates at $\sim 55\, \rm{\micron}$ \citep{koi03}, fullerenes (C$_{60}$ and C$_{70}$), and PAHs (see the companion paper by \citealt{kan17} in this issue).

So far, the use of IR tracers based on mid- to far-IR spectroscopic observations with \textit{Spitzer} and \textit{Herschel} to study the chemical evolution of galaxies was restricted only to very luminous sources, due to sensitivity issues \citep[e.g.][]{rui13,bri15}. Future facilities planned for the next decade will not cover this key spectral range. Therefore their study of the chemical evolution in galaxies will be limited to unobscured regions in galaxies, and will rely on tracers with a strong dependency on the gas temperature (see Section\,\ref{syner}).

The unique sensitivity of an IR observatory with the characteristics of \textit{SPICA} would enable the study of key processes in the baryon cycle that are hidden to optical wavelengths. A description of the \textit{SPICA} spectrometers, SMI and SAFARI, is given in \citet{roe17}.
Fig.\,\ref{simul} shows the predicted spectra for starburst galaxies and AGN out to $z = 4$, assuming the spectral characteristics of a local starburst galaxy, and an AGN. The simulated spectra in this figure are based on the \textit{ISO}/SWS + LWS and \textit{Spitzer}/IRS observations of the starburst galaxy M82 and the Seyfert 2 galaxy NGC\,1068. These galaxies were selected due to the availability of high S/N \textit{ISO}-SWS and -LWS spectra which, combined with the high-spectral resolution \textit{Spitzer}/IRS observations, allowed us to build spectral templates with a wide coverage in wavelength from $10$ to $250\, \rm{\micron}$, similar to the rest-frame range that would be accessible to \textit{SPICA}. The spectra of both galaxies were scaled to a luminosity of $2 \times 10^{12}\, \rm{L_\odot}$, and located at $z = 1, 2, 3,$ and $4$ (from top to bottom). The different observational modes of SMI and SAFARI would be sensitive enough to provide access to a unique suite of IR lines in the mid- to far-IR rest-frame spectra of starburst galaxies and AGN at the knee of the luminosity function up to $z \sim 3$.

In the following sections we will focus on the observational strategies that \textit{SPICA} could implement to determine heavy element abundances of galaxies from high-$z$ to the Local Universe. First, we will describe the direct abundance determinations in Section\,\ref{direct}, followed by the indirect methods in Sections\,\ref{o3n3} and \ref{ne23s34}. The main diagnostics are summarised in Table\,\ref{tabdiag}.
\begin{table}
\caption{Direct (upper rows) and indirect (lower rows) metallicity tracers based on the mid- to far-IR lines that could be exploited by \textit{SPICA}. In the table, ``[X]'' refers to any of the fine-structure lines found in the range accessible to \textit{SPICA} (e.g. [\ion{S}{iii}]$_{18.7 \rm{\micron}}$, [\ion{Ar}{ii}]$_{6.99 \rm{\micron}}$, [\ion{Ne}{ii}]$_{12.8 \rm {\micron}}$) whose corresponding ionic stage abundance can be determined from the flux ratio of the [X] line to any of the hydrogen recombination lines. The Hu$\alpha$ line is relatively weak and thus is expected to be detected only in nearby galaxies ($z \sim 0$).}
\begin{center}
\normalsize
\begin{tabular}{@{}lcc@{}}
\hline\hline
 Type & Redshift range & Tracer\\
\hline\\[-0.3cm]
 Direct &  Local    & $\frac{[\rm{X}]}{\rm{Hu\alpha}}$ \\[0.2cm]
  &  1.5--3   & $\frac{[\rm{X}]}{\rm{Pfund\alpha}}$ \\[0.2cm]
\hline\\[-0.3cm]
 Indirect & 0--1.6   & $\frac{2.2 \times [\rm{OIII}]88 \rm{\micron} \, + \, [\rm{OIII}]52 \rm{\micron}}{[\rm{NIII}]57 \rm{\micron}}$ \\[0.2cm]
 & 0--3.0   & $\frac{[\rm{OIII}]52 \rm{\micron}}{[\rm{NIII}]57 \rm{\micron}}$ \\[0.2cm]
 & 0.15--3  & $\frac{[\rm{NeIII}]12.8 \rm{\micron} \, + \, [\rm{NeIII}]15.6 \rm{\micron}}{[\rm{SIII}]18.7 \rm{\micron} \, + \, [\rm{SIV}]10.5 \rm{\micron}}$ \\[0.2cm]
\hline\hline
\end{tabular}
\end{center}
\label{tabdiag}
\end{table}

\subsection{Direct Abundance Determinations}\label{direct}
A direct determination for the abundances of the main ionic stages of several heavy elements could be obtained using ratios of IR fine-structure lines to hydrogen recombination lines. In the case of \textit{SPICA}, this would be possible using Pfund-$\alpha$ --\,at $7.46\, \rm{\micron}$ in the rest frame\,-- or Humphreys-$\alpha$ --\,\ion{H}{i}\,(7-6) or Hu$\alpha$ at $12.37\, \rm{\micron}$. At $1.5 < z \lesssim 3$, the medium-resolution mode (MR, $R \sim 1500$) of SMI would be able to cover simultaneously Pfund-$\alpha$ and the argon lines [\ion{Ar}{ii}]$_{6.99 \rm{\micron}}$ and [\ion{Ar}{iii}]$_{8.99 \rm{\micron}}$. For instance, Pfund-$\alpha$ could be detected by SMI/MR in a starburst galaxy with similar spectral characteristics to M82, scaled to $2 \times 10^{12}\, \rm{L_\odot}$, as shown by the inset in Fig.\,\ref{simul} (left panel, at $z = 3$). Then, the gas metallicity could be estimated from the ratio of argon to hydrogen lines, since argon does not suffer from depletion \citep{ver03}. Since the ionisation potentials of Ar$^0$ and Ar$^{2+}$ are $15.8\, \rm{eV}$ ($> 13.6\, \rm{eV}$) and $40.7\, \rm{eV}$ ($< 54.4\, \rm{eV}$), respectively, the presence of Ar$^0$ and Ar$^{3+}$ could be expected. In the case of neutral argon, observations of star forming regions do not show a significant contribution from this species, suggesting that most of the argon in the nebula is ionised \citep[e.g.][]{leb04,leb06}. This could be explained by a high photoionisation cross section in the neutral argon \citep{sof98}. On the other hand, Ar$^{2+}$ has a relatively low photoionisation cross-section for energies below $70\, \rm{eV}$ \citep{ver96}, and thus Ar$^{3+}$ has a minor contribution for most \ion{H}{ii} regions \citep[$\lesssim 3\%$, e.g.][]{pei03,tsa03,g-r06,leb08}, although this could increase significantly when exposed to harder radiation fields in low-metallicity environments \citep[$\sim 20\%$,][]{pei00}. Therefore, the total argon abundance could be derived using an ionisation correction factor to account for the contribution of Ar$^{3+}$ \citep[e.g.][]{ver02,per07,hag08}.

In combination with the low-resolution mode (LR, $R \sim 300$) of SAFARI, it would be also possible to obtain a direct determination of the neon abundance for star forming galaxies up to $z \sim 3$, based on the [\ion{Ne}{ii}]$_{12.8 \rm{\micron}}$ and [\ion{Ne}{iii}]$_{15.6 \rm{\micron}}$ lines. Ne$^{2+}$ has a high ionisation potential ($63.5\, \rm{eV}$) above the helium absorption edge in stars ($54.4\, \rm{eV}$), and consequently Ne$^{3+}$ does not exist in significant amounts \citep{ber09b,hag08,dor13}. At the lower end, neutral neon could exist within the Str\"omgren radius due to its relatively high ionisation potential ($21.6\, \rm{eV}$). However, a further investigation using photoionisation models proved that the expected amount of Ne$^0$ within the nebula is negligible, with an ionisation correction factor below $\lesssim 1.05$ for a wide range of ionisation parameter values \citep{mar02}. In the case of sulphur, a minor contribution of S$^{4+}$ is expected due to the high ionisation potential of S$^{3+}$ ($47.3\, \rm{eV}$), however S$^+$ has an ionisation potential of $23.3\, \rm{eV}$ and thus it could be present in the nebula. Based on measurements of the sulphur abundance in nearby star forming regions \citep{pei03,ver02,g-r06} and photoionisation models \citep{tsa05}, the expected amount of S$^+$ is less than $\sim 10\%$ of the total sulphur abundance \citep{leb08,ber09b}. As in the case of argon, the gas-phase abundance of sulphur could be derived from the [\ion{S}{iii}]$_{18.7 \rm{\micron}}$ and [\ion{S}{iv}]$_{10.5 \rm{\micron}}$ lines and the hydrogen recombination lines detected by SMI and SAFARI, using a relatively small ionisation correction factor.

In nearby galaxies, direct abundances could be determined using the Hu$\alpha$ line (e.g. \mbox{\citealt{ber09b}}), covered by SMI at high-spectral resolution (HR, $R \simeq 28\,000$). As mentioned earlier, the weak dependence of the IR lines with the density and the effect of the radiation field hardness could be constrained using the corresponding diagnostics based on ratios of mid- and far-IR lines \citep[e.g.][]{cro13}. Besides argon, neon, and sulphur, we would be able to provide abundances for the dominant ionic species of nitrogen (N$^+$ and N$^{2+}$) and help in the determination of direct abundances for oxygen (O$^0$, O$^{2+}$, and O$^{3+}$), iron (Fe$^+$ and Fe$^{2+}$), and carbon (C$^+$), based on the spectral range covered by SMI and SAFARI ([\ion{N}{ii}]$_{122 \rm{\micron}}$ and [\ion{N}{iii}]$_{57 \rm{\micron}}$; [\ion{O}{i}]$_{63, 145 \rm{\micron}}$, [\ion{O}{iii}]$_{52, 88 \rm{\micron}}$, and [\ion{O}{iv}]$_{25.9 \rm{\micron}}$; [\ion{Fe}{ii}]$_{17.9, 26.0 \rm{\micron}}$ and [\ion{Fe}{iii}]$_{22.9 \rm{\micron}}$; [\ion{C}{ii}]$_{158 \rm{\micron}}$).

\subsection{The O$_3$N$_3$ index}\label{o3n3}
\begin{figure}[h!!!!!]
\begin{center}
  \includegraphics[width=\columnwidth]{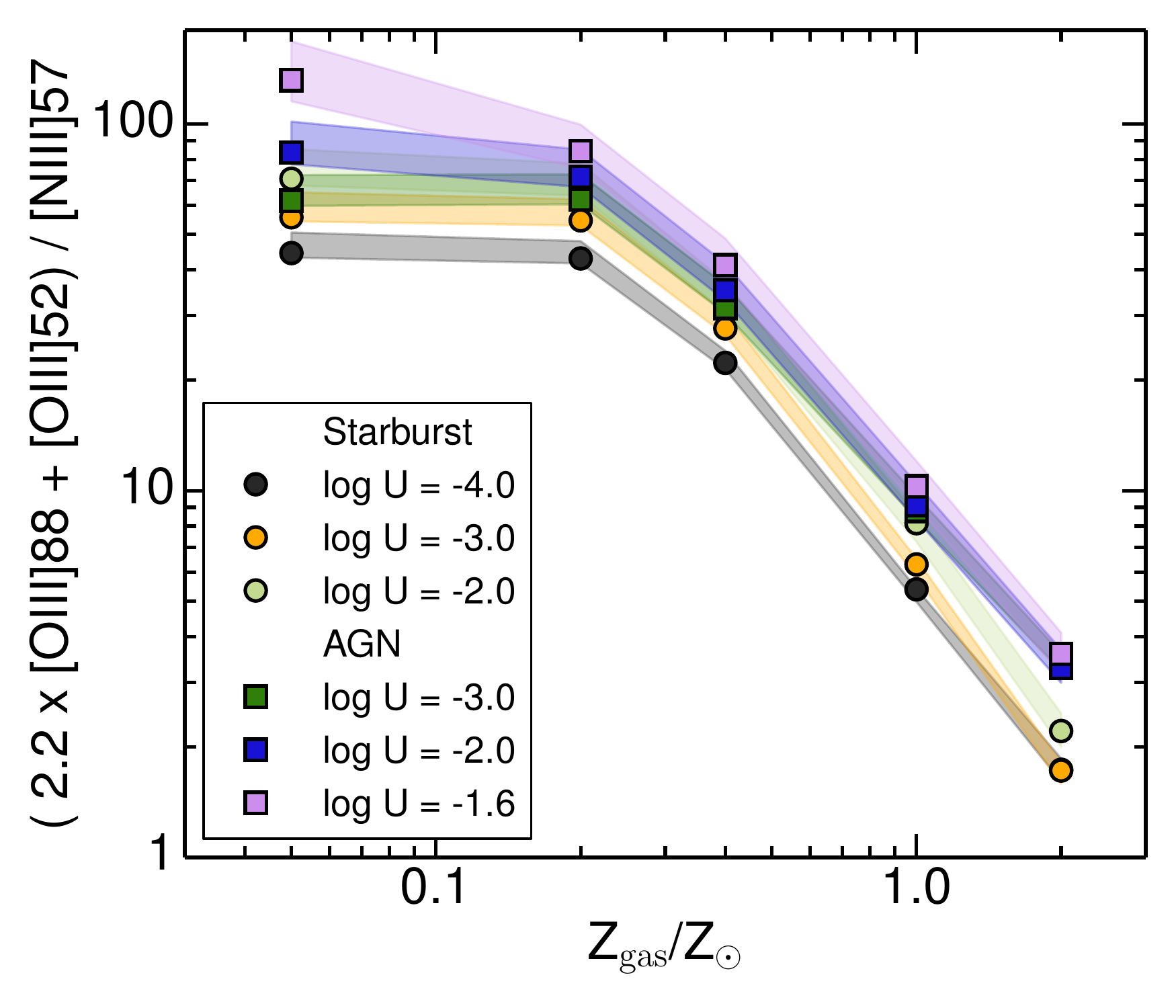}
  \includegraphics[width=\columnwidth]{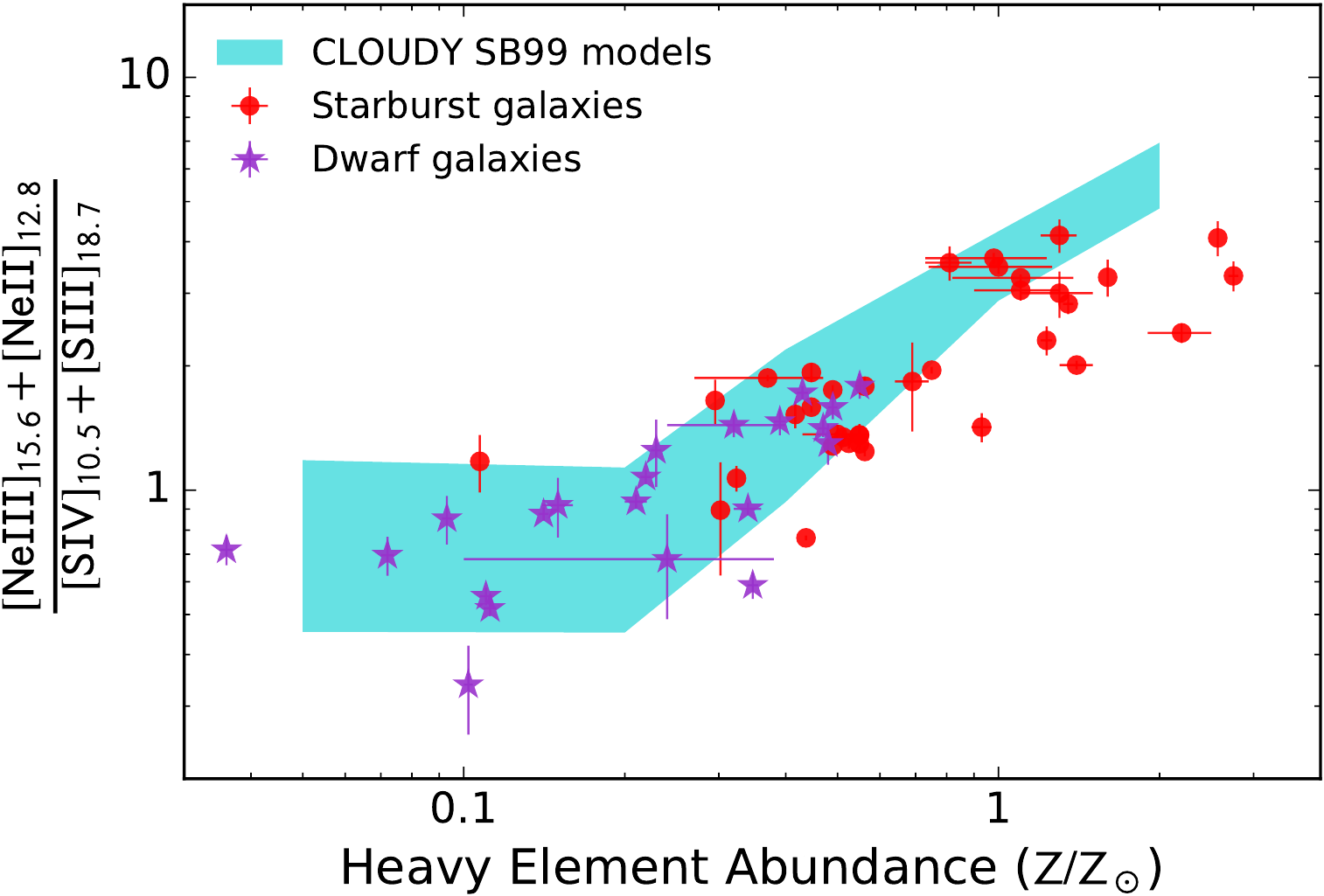}
  \caption{\textit{Top:} AGN and starburst models for the metallicity sensitive ($2.2 \times$[\ion{O}{iii}]$_{88 \rm{\micron}}$+[\ion{O}{iii}]$_{52 \rm{\micron}}$)/[\ion{N}{iii}]$_{57 \rm{\micron}}$ line ratio as a function of the gas-phase metallicity. For each metallicity bin, those models with the same ionisation parameter but different densities are grouped and their median ratio is indicated by a circle (square) for the starburst (AGN) models. The shaded area represents the scatter due to the gas density dependence of this ratio for a given ionisation parameter. Figure from \citet{per17}. \textit{Bottom:} the ([\ion{Ne}{ii}]$_{12.8 \rm{\micron}}$ + [\ion{Ne}{iii}]$_{15.6 \rm{\micron}}$) to ([\ion{S}{iii}]$_{18.7 \rm{\micron}}$ + [\ion{S}{iv}]$_{10.5 \rm{\micron}}$) line ratio (i.e. the Ne$_{23}$S$_{34}$ index) from \textit{Spitzer}/IRS observations of starburst galaxies in the Local Universe vs. indirect gas-phase metallicity determined from strong optical lines \citep{mou10,pil14}. The \textsc{Cloudy} simulations including sulphur stagnation above $Z > 1/5\, \rm{Z_\odot}$ are in agreement with the observed increase of the line ratio with increasing metallicity for $Z \gtrsim 0.2\, \rm{Z_\odot}$ \citep[figure adapted from][]{fo16}. Below $\lesssim 0.2\, \rm{Z_\odot}$, direct abundances can be estimated using, e.g. the ratio of [\ion{Ar}{ii}]$_{6.99 \rm{\micron}}$ and [\ion{Ar}{iii}]$_{8.99 \rm{\micron}}$ to the Pfund-$\alpha$ line for galaxies at $1.5 < z \lesssim 3$ (see Section\,\ref{direct}).}\label{diag}
\end{center}
\end{figure}

One of the main IR diagnostics to measure the metallicity of galaxies is based on the ratio of the [\ion{O}{iii}]$_{52, 88 \rm{\micron}}$ to [\ion{N}{iii}]$_{57 \rm{\micron}}$ lines. These are two key elements in the gas chemistry: oxygen is an alpha-capture primary element produced by massive stars, whereas nitrogen behaves as both a primary and a secondary nucleosynthetic element. As a result, the N/O relative abundance shows a dependence with the star-formation history of the galaxy, and the subsequent global metallicity \citep{lia06,pil14,vin16}, which could be determined by \textit{SPICA} 
(\citealt{nag11}; \citealt{per17}). O and N have similar ionisation potentials for their first five ionisation stages and, thus, both have very similar ionisation structures almost independent of the ionisation parameter and hardness of the radiation field. Therefore, although ratios of [\ion{O}{iii}] to [\ion{N}{iii}] lines only trace the relative abundance of O$^{2+}$ to N$^{2+}$, this ratio serves as a very good proxy for the global N/O abundance ratio (see \citealt{p-c09} for a calibration using the optical [\ion{N}{ii}]$\,\lambda 6584$/[\ion{O}{ii}]$\,\lambda 3727$ line ratio. The [\ion{O}{iii}] lines at $52$ and $88\, \rm{\micron}$, whose ratio can be used as a density tracer for the ionised gas in the $\sim 10^2$--$10^3\, \rm{cm^{-3}}$ range, are combined to produce a tracer for the gas metallicity with a weak dependency on the density. This tracer is based on the ratio of ($2.2 \times$[\ion{O}{iii}]$_{88 \rm{\micron}}$\,+\,[\ion{O}{iii}]$_{52 \rm{\micron}}$) to [\ion{N}{iii}]$_{57 \rm{\micron}}$ (hereafter the O$_3$N$_3$ index), as shown by the top panel in Fig.\,\ref{diag}, it can be applied to both starburst galaxies and AGN, almost unaffected by the temperature, the extinction, the density, and the hardness of the radiation field (for further details see \citealt{per17}).

Gas metallicities can be determined using the diagnostic shown in Fig.\,\ref{diag} up to $z \sim 1.6$, where the [\ion{O}{iii}]$_{88 \rm{\micron}}$ line would still fall within the spectral range of SAFARI ($35$--$230\, \rm{\micron}$). A diagnostic based on the [\ion{O}{iii}]$_{52 \rm{\micron}}$/[\ion{N}{iii}]$_{57 \rm{\micron}}$ ratio \citep[see][]{nag11} is also a metallicity tracer up to $z \sim 3$ if the density is constrained through other line ratios, e.g. from the [\ion{S}{iii}]$_{18.7, 33.5 \rm{\micron}}$ or the [\ion{Ne}{iii}]$_{15.6, 36.0 \rm{\micron}}$ lines. As shown in Fig.\,\ref{simul}, a telescope like \textit{SPICA} would be sensitive enough to detect these lines in a few hours of observing time in typical starburst galaxies and AGN with luminosities close to the knee of the luminosity function at $z \sim 3$.

\subsection{The Ne$_{23}$S$_{34}$ index}\label{ne23s34}
Above $z \gtrsim 0.15$ and $0.7$, the [\ion{S}{iv}]$_{10.5 \rm{\micron}}$ line would enter in the SMI/HR and MR ranges, respectively, enabling an additional indirect abundance diagnostic --\,based on the calibration of metallicity-sensitive line ratios\,-- using the ([\ion{Ne}{ii}]$_{12.8 \rm{\micron}}$ + [\ion{Ne}{iii}]$_{15.6 \rm{\micron}}$) to ([\ion{S}{iii}]$_{18.7 \rm{\micron}}$ + [\ion{S}{iv}]$_{10.5 \rm{\micron}}$) line ratio (hereafter the Ne$_{23}$S$_{34}$ index; lower panel in Fig.\,\ref{diag}; see \mbox{\citealt{fo16}} for further details). The chemical species of Ne$^+$ and Ne$^{2+}$ on one hand, and S$^{2+}$ and S$^{3+}$ on the other, are the most important stages of ionisation for neon and sulphur, respectively, in starburst galaxies and AGN. However, the stagnation of the sulphur abundance, possibly caused by depletion of sulphur onto dust grains \citep{ver03,dor16,vid17}, prevent the rise of the sulphur line intensities with increasing metallicity. Thus, the total [\ion{S}{iv}]$_{10.5 \rm{\micron}}$ + [\ion{S}{iii}]$_{18.7 \rm{\micron}}$ remains almost constant, while the [\ion{Ne}{iii}]$_{15.6 \rm{\micron}}$ + [\ion{Ne}{ii}]$_{12.8 \rm{\micron}}$ increases with the neon abundance, which is not depleted onto dust grains.

The bottom panel in Fig.\,\ref{diag} shows a calibration of this metallicity sensitive ratio obtained from \textit{Spitzer}/IRS spectra in the high-resolution mode ($R \sim 600$) vs. gas-metallicities obtained from indirect optical tracers for a sample of unobscured local starburst and dwarf galaxies (\mbox{\citealt{mou10}}, \mbox{\citealt{pil14}}). The blue shaded area corresponds to photoionisation models obtained with \textsc{Cloudy} simulations \citep{fer17} for a starburst galaxy, assuming a constant sulphur abundance above $Z = 0.2\, \rm{Z_\odot}$, while the neon abundance increases linearly with the overall gas metallicity (see \citealt{fo16} for a full description of these models). Both the theoretical predictions and the measurements of unobscured starburst and dwarf galaxies are in agreement, since the trend shown by the models with sulphur abundance stagnation reproduces the behaviour of the observed line ratios. Since [\ion{Ne}{iii}] and [\ion{Ne}{ii}] are the most important stages of ionisation, the neon abundances determined with this method will be almost independent of ionisation correction factors.

All the lines used in this diagnostic would be covered by SMI and SAFARI for galaxies at $0.15 < z < 3$. The weakest line in starburst galaxies with solar metallicities is [\ion{S}{iv}]$_{10.5 \rm{\micron}}$ (left panel in Fig.\,\ref{simul}), although it is expected to become stronger with decresing metallicity, as shown by the mid-IR spectra of low-metallicity dwarf galaxies in the Local Universe (e.g. \mbox{\citealt{mad06}}, \mbox{\citealt{cor15}}).
Therefore, high-$z$ galaxies with sub-solar metallicities are expected to show stronger [\ion{S}{iv}]$_{10.5 \rm{\micron}}$ relative to other mid-IR fine-structure lines when compared to the spectrum of the starburst galaxy in Fig.\,\ref{simul}. A different depletion pattern as that implied by Fig.\,\ref{diag} is not expected at high-$z$, since the range of metallicities probed by this diagram covers the range of values expected for galaxies at $z \lesssim 4$ (e.g. \citealt{dav17}). Still, this scenario could be explored using observations of the dust spectral features in the mid- to far-IR spectra of galaxies at high-$z$ (see Section\,\ref{spicadust}). At very low metallicities, studies based on galaxies in the Local Universe predict a constant N/O abundance ratio due to the primary origin of nitrogen in massive stars \citep{zee98}, while sulphur does not show a significant abundance stagnation \citep{dor16}. Thus, both indirect tracers O$_3$N$_3$ and Ne$_{23}$S$_{34}$ show a flat behaviour below $\lesssim 0.2\, \rm{Z_\odot}$ in Fig.\,\ref{diag}. \textit{SPICA} would be able to probe if this scenario holds at high-$z$. For instance, enhanced N/O abundances have been reported at $z \sim 2$ with regard to those of Local galaxies at similar gas-phase metallicities, possibly due to the high specific SFR at high-$z$ (see discussion in \citealt{mas14,mas16}). On the other hand, direct abundance estimates for several ionic species can be obtained for neon, sulphur, oxygen, nitrogen, iron, and carbon below $0.2\, \rm{Z_\odot}$ in galaxies at $1.5 < z \lesssim 3$, from ratios of the mid- and far-IR fine-structure lines of these elements to the Pfund-$\alpha$ line, covered by SMI/MR and SAFARI/LR (see Section\,\ref{direct}; Fig.\,\ref{simul}).

\section{\textsc{SPICA} AND THE BUILD-UP OF METALS AND DUST}\label{buildup}

We discuss here the role that \textit{SPICA} would play to address the science topics described in Section\,\ref{chevol} --\,i.e. the assembly of galaxies, the evolution of metallicity in galaxies, and the buil-up of cosmic dust\,-- using the tracers for heavy element abundances and dust composition described in Section\,\ref{observables}.

\subsection{The Assembly of Galaxies}\label{spicanear}
The potential of an observatory like \textit{SPICA} to provide high-sensitivity IR spectral mapping of galaxies in the Local Universe, able to peer into their most obscured regions, would play an essential role to understand how galaxies were assembled. In the low-resolution mode, \textit{SPICA} would be an extremely efficient mapping machine, reaching line luminosity noise levels of a few $10^{-19}\, \rm{W\,m^{-2}}$ over a square arcmin in less than a minute of observing time and over the entire wavelength range. Moreover, the $2.5\, \rm{m}$ mirror of \textit{SPICA} would provide a diffraction-limited angular resolution of $\sim 2''$ to $32''$ in the $20$--$350\, \rm{\micron}$ range, which corresponds to $100\, \rm{pc}$--$1.5\, \rm{kpc}$ at $10\, \rm{Mpc}$, respectively. For example, this angular resolution would be sufficient to resolve individual molecular clouds throughout galaxies in the Local Group (Milky Way, M31, M33, IC10, LMC, NGC\,6822, SMC, and WLM, in order of decreasing metallicity).
Metallicity determinations based on the O$_3$N$_3$ index could be obtained at $\sim 7''$ resolution ($\sim 0.3\, \rm{kpc}$ at $10\, \rm{Mpc}$), i.e. a fraction of about 1/100 of the diameter for a galaxy located at $10\, \rm{Mpc}$ with the size of the Milky Way. \textit{SPICA} would be able to map a typical nearby spiral galaxy ($\sim 25\, \rm{arcmin^2}$) with both SMI and SAFARI at medium to high-spectral resolution in about $\sim 20$ hours \citep[see][]{tak17}.

The sensitivity and efficiency of an observatory like \textit{SPICA} would allow us to quickly map large areas of nearby objects at the diffraction limit of the $2.5\, \rm{m}$ mirror above $\gtrsim 20\, \rm{\micron}$. For instance, when compared with the \textit{Herschel}/PACS spectral maps for NGC\,891 \citep[at $9.6\, \rm{Mpc}$][]{hug15}, \textit{SPICA} would provide: \textit{i)} about 100 times better sensitivity, thus revealing the outer galaxy regions with lower-surface brightness ($\sim 10^{-11}\, \rm{W\,m^{-2}\,sr^{-1}}$, $5 \sigma$, $1\, \rm{hr.}$ compared to $\sim 5 \times 10^{-9}\, \rm{W\,m^{-2}\,sr^{-1}}$ for the [\ion{O}{iii}]$_ {88 \rm{\micron}}$ line with PACS); \textit{ii)} better spatial sampling ($\sim 0.5\, \rm{kpc}$/pixel in PACS); and \textit{iii)} with a much larger efficiency to cover the full extension of NGC\,891 ($\sim 7.7\, \rm{arcmin^2}$) for the whole $12$--$230\, \rm{\micron}$ spectral range in about $8\, \rm{hr.}$ --\,PACS measured small spectral windows centred on the [\ion{O}{i}]$_{63, 145 \rm{\micron}}$, [\ion{O}{iii}]$_{88 \rm{\micron}}$, [\ion{N}{ii}]$_{122 \rm{\micron}}$, and [\ion{C}{ii}]$_{158 \rm{\micron}}$ lines, covering half of this galaxy in a total of $\sim 12$ hours.

These capabilities would allow \textit{SPICA} to trace the spatial distribution of heavy elements in ISM of nearby galaxies. The metal budget at different radii would be compared to the stellar metallicities, which can be derived from optical and near-IR spectroscopic studies \citep[e.g.][]{coc13,bel16}, and chemical evolution models. This approach would allow to estimate the average outflow loading factors over the galaxies lifetime --\,i.e. the ratio between the star formation and the mass outflow rates\,-- which is an essential step to constrain the episodes of matter cycling, and ultimately to understand how nearby galaxies were assembled \citep[see][]{bel16,gon17}. Therefore, a detailed study of the spatial distribution of heavy element abundances in nearby galaxies would be a unique probe of the past history in the evolution of these galaxies.


\subsection{The Evolution of Metallicity in Galaxies}\label{spicaevol}
\textit{SPICA} would be able to trace the global content of metals in galaxies from $z \sim 3$ till the present time.
To accomplish this, future studies based on \textit{SPICA} observations would dispose of heavy element abundances for galaxies up to $z \sim 3$, obtained from tracers with a feeble response to both extinction and temperature: e.g. using direct measurements of the main ionic species based on ratios of IR fine-structure lines to \mbox{Pfund-$\alpha$} or Hu$\alpha$ and ionisation correction factors (see Section\,\ref{direct}; Fig.\,\ref{simul}); and/or indirect methods based on ratios of strong IR fine-structure lines such as those involved in the O$_3$N$_3$ and the Ne$_{23}$S$_{34}$ indices (Sections\,\ref{o3n3} and \ref{ne23s34}, respectively; Fig.\,\ref{diag}). A spectroscopic survey designed to observe thousand of galaxies at $z < 3$ --\,e.g. covering a wide range in luminosity, galaxy type\,-- would allow to obtain a statistically valid assessment of the metallicity evolution in galaxies. \textit{SPICA} would be able to accomplish such a survey in $\sim 4\,000$ hours of observing time.


Once the metallicities are measured using IR tracers, a global view of the baryon cycle could be derived by including the two remaining ingredients, i.e. the stellar mass and the total gas content of these galaxies. The latter would rely mainly on future and present facilities. For instance, the stellar mass can be obtained from near- and mid-IR photometric surveys available nowadays (e.g. with \textit{AKARI}, \citealt{ish10}; \textit{WISE}, \citealt{wri10}; \textit{Spitzer}, \citealt{ash13}), plus future catalogues assembled by the extremely-large telescopes (ELTs) --\,the European-ELT\footnote{\url{http://www.eso.org/sci/facilities/eelt/}}, the Thirty Meter Telescope \citep{san13}, and the Giant Magellan Telescope \citep{she10}, \textit{Euclid} \citep{mac16}, the \textit{Wide Field Infrared Survey Telescope} \citep[\textit{WFIRST};][]{spe13}, and available \textit{James Webb Space Telescope} \citep[\textit{JWST};][]{gar06} observations at the time. On the other hand, the molecular gas content can be derived from follow-up observations of these sources with the Atacama Large Millimeter/submillimeter Array (ALMA; \mbox{\citealt{woo09}}) and the NOrthern Extended Millimeter Array (NOEMA\footnote{\url{http://www.iram-institute.org/EN/noema-project.php}}) in the CO and [\ion{C}{i}]$_{371, 609 \rm{\micron}}$ transitions, assuming a conversion factor\footnote{The dependence of this conversion factor with metallicity \citep[e.g.][]{sch12} could be calibrated for nearby galaxies using both the HD lines at $56$ and $112\, \rm{\micron}$ and the [\ion{C}{ii}]$_{158 \rm{\micron}}$ line to detect CO-dark gas \citep[see][]{tak17}. \textit{SPICA} would detect the HD lines in typical galaxies up to $\sim 10\, \rm{Mpc}$.} from CO or \ion{C}{i} to H$_{\rm 2}$ \citep[e.g.][]{pap04}. Alternatively, a proxy for the molecular gas content could be inferred from the \textit{SPICA} spectra using IR transitions sensitive to the SFR such as the neon lines \citep{ho07}, or from the modelling of the $24$--$60\, \rm{\micron}$ rest-frame continuum which is also sensitive to the SFR \citep{cie14,tak17}. To remove the possible AGN contamination, the relative AGN/starburst contribution could be derived from the ratio of [\ion{O}{iv}]$_{25.9 \rm{\micron}}$ to either [\ion{O}{iii}]$_{52, 88 \rm{\micron}}$ or the total [\ion{Ne}{ii}]$_{12.8 \rm{\micron}}$ + [\ion{Ne}{iii}]$_{15.6 \rm{\micron}}$ \citep{spi15,fo16}. Finally, the atomic gas content could be estimated from observations in the \ion{H}{i} line, e.g. using the Square Kilometer Array \mbox{\citep[SKA;][]{dew09}}, which will be already operational by the mid-2020s, though fully developed at the end of the decade.

\subsection{The Build-Up of Cosmic Dust}\label{spicadust}
The unique sensitivity of a cryogenic telescope with the characteristics of \textit{SPICA} would enable the study of the dust composition based on several spectral features in the mid- to far-IR range (Section\,\ref{whyIR}; \citealt{tak17}), largely improving the work based on \textit{Spitzer} and \textit{Herschel} spectroscopic observations \citep[e.g.][]{spo06,mat15}. A whole new suite of solid-state features can be used to investigate the main sources of dust in the ISM of galaxies at high-$z$. Among the different sources that contribute to the dust reservoir in galaxies one can find: evolved stars, producing crystalline silicates \citep{mol99,mol02}, dolomite and calcite \citep{kem02}, and complex carbon molecules such as C$_{60}$ and C$_{70}$ \citep{cam10}; post-AGB stars show a rich IR spectrum, as shown by \citet{mat14}, with features at $15.8\, \rm{\micron}$ and $21\, \rm{\micron}$ (associated with PAHs), and $30\, \rm{\micron}$ (ascribed to MgS); supernovae remnants (SNR) may also be a significant source of dust at high-$z$, e.g. Cassiopeia\,A shows a prominent $21\, \rm{\micron}$ feature associated with FeO and possibly SiO$_2$ \citep{rho08}.
\begin{figure}
  \begin{center}
  \includegraphics[width=\columnwidth]{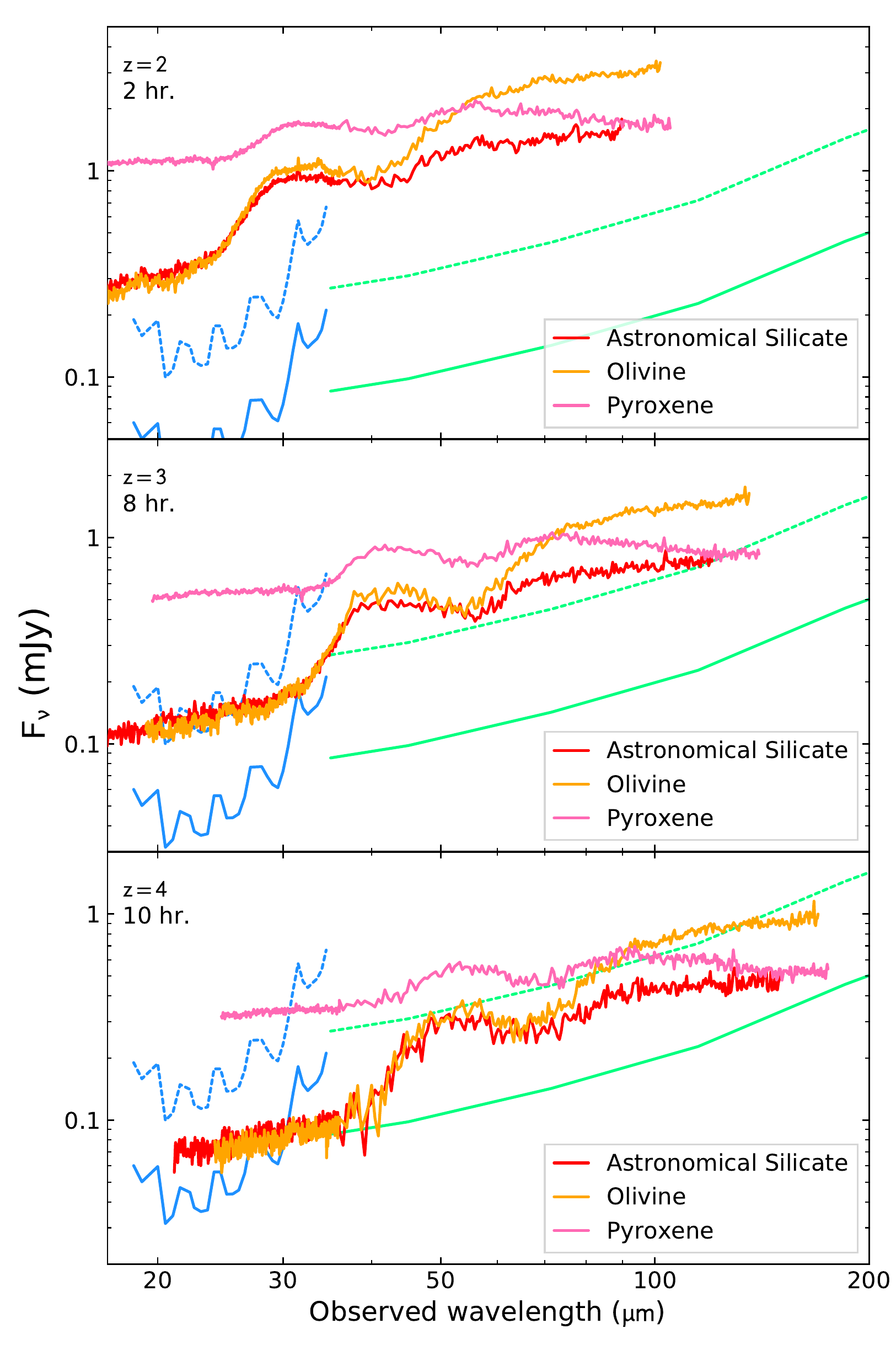}
  \caption{Differences in the \textit{Spitzer}/IRS spectrum of three quasars associated to different silicate compositions \citep[see][]{xie17}: PG\,1004+130 (astronomical silicate, in red colour), PG\,1351+640 (amorphous olivine, in yellow), and PG\,2214+139 (amorphous pyroxene, in pink). In this simulation the spectra have been scaled to a luminosity of $10^{12}\, \rm{L_\odot}$, and located at $z = 2$ (top panel, 2 hours of integration time), $z = 3$ (central panel, 8 hours), and $z = 4$ (bottom panel). The \textit{SPICA} $5 \sigma$ sensitivities for SMI/MR (in blue) and SAFARI/LR (in green) are indicated for integration times of 1 hour (dotted lines) and 10 hours (solid lines).}\label{dustfig}
\end{center}
\end{figure}

In nearby galaxies, medium and high-resolution spectroscopy with SMI and SAFARI would be able to resolve the emission of SNR from the surrounding ISM at $\sim 1\, \rm{Mpc}$ (assuming a SNR size of $\sim 60\, \rm{pc}$; \citealt{elw17}) and investigate the properties and composition of the surviving dust after the pass of the reverse shock. Only far-IR follow-up observations of SNR in the Local Universe can probe the dust composition and size distribution produced by these sources. Specifically, future observations of SN\,1987A would be of particular interest to constrain current models of dust condensation efficiency after the passage of the reverse shock. A detailed discussion on the role of \textit{SPICA} in the study of dust properties in nearby galaxies can be found in the companion paper by \citet{tak17}.


Focusing on studies at high-$z$, a cryogenic IR observatory like \textit{SPICA} would be sensitive enough to distinguish among different shapes in the rest-frame mid-IR spectrum of galaxies associated with differences in the chemical composition of dust. The predicted shape of the mid- to far-IR spectra for three quasars with different dust compositions, from the results of \citet{xie17}, are shown in Fig.\,\ref{dustfig} scaled to a luminosity of $2 \times 10^{12}\, \rm{L_\odot}$. The dust composition determines the central wavelength, width, and relative intensity of the two silicate bands at $9.7\, \rm{\micron}$ and $18\, \rm{\micron}$. Thus, a detailed modelling of the continuum emission and the dust emission and absorption features in this range is able to probe the dust composition \citep[e.g.][]{spo06}. The quasar spectra shown in Fig.\,\ref{dustfig} are dominated by astronomical silicate (PG\,1004+130, in red), amorphous olivine (PG\,1351+640, in yellow), and amorphous pyroxene (PG\,2214+139, in pink). Therefore, the access to dust features in the rest-frame mid- to far-IR range with \textit{SPICA} would allow us to explore, for the first time, the composition of dust in galaxies at the peak of the star formation and SMBH accretion activity ($1 < z < 3$). At farther distances, $z \gtrsim 4$, very little is known on the dust produced by metal-free SNe \citep{noz03,sch04}, which could be investigated through mid-IR bands such as those produced by magnetite (Fe$_3$O$_4$), as discussed in a forthcoming study by Egami et al. (in preparation).

Finally, other agents involved in the baryon cycle can have a strong influence in the dust content and composition. For instance, outflows originated by AGN and star formation activity can remove dust from the ISM, while inflows from the halo and the IGM surrounding the galaxy can alter the dust content and composition \citep{cie15,gon17}. Only an IR observatory with the characteristics of \textit{SPICA} could determine the impact of these processes in the ISM chemistry.




\section{Synergies with Future Facilities}\label{syner}

\textit{SPICA} is planned for the late 2020s, during a lively period for multi-wavelength astronomy and galaxy evolution studies in particular.
However, the observations that would be provided by \textit{SPICA} are not accessible to any of the future generation of ground-based ELTs nor to \textit{JWST}. These facilities will have access to the rest-frame optical and near-IR spectra of high-$z$ galaxies ($z \gtrsim 1$) with a limited access to the mid-IR range ($\lambda \lesssim 7\, \rm{\micron}$ at $z = 3$). Therefore, the use of IR tracers based on \textit{JWST} and ELTs observations will be restricted to galaxies in the Local and nearby ($z < 0.5$) Universe.

At the time \textit{SPICA} will fly, the catalogues delivered by \textit{Euclid} and \textit{WFIRST} in the near-IR, by the \textit{Advanced Telescope for High ENergy Astrophysics} \citep[\textit{Athena}+;][]{nan13} in the X-ray range, will be already available, containing a vast number of sources at $1 < z < 3$. Together with the galaxies that would be eventually discovered by a spectrophotometric survey with SMI in the mid-IR \citep[see][]{gru17}, these catalogues would be useful to define a sample of galaxies to study the chemical evolution over cosmic time. This sample would be observed with mid- to high resolution spectroscopy using SMI and SAFARI.
From these catalogues, \textit{SPICA} would be able to identify candidates to perform follow-up spectroscopic observations in the mid- to far-IR rest-frame range with medium resolution ($R \gtrsim 300$) SAFARI and SMI. These observations could be the basis for the study of the chemical evolution of galaxies (see Section\,\ref{spicaevol}). Of particular interest in this field are the synergies of SMI and SAFARI spectroscopy with the Mid-InfraRed Instrument (MIRI) on-board \textit{JWST}. The latter could provide abundance determinations for the main ionic species of several elements (see Section\,\ref{whyIR}) in galaxies at $1 < z \lesssim 3$ based on the combination of Br$\alpha$ ($4.051\, \rm{\micron}$ in the rest-frame), Pa$\alpha$ ($1.875\, \rm{\micron}$), and Pa$\beta$ ($1.282\, \rm{\micron}$) with \textit{JWST} and the fine-structure lines of the chemical species in the mid- to far-IR range provided by \textit{SPICA}. On the other hand, far-IR lines such as [\ion{N}{ii}]$_{57 \rm{\micron}}$ and [\ion{N}{iii}]$_{122 \rm{\micron}}$ would provide an abundance estimate for the main ionisation stages of nitrogen, when coupled with the hydrogen recombination lines in the near-IR provided by \textit{JWST}.

The synergy between \textit{SPICA} and \textit{Athena}+ would allow us to address the problem of the ``missing metals'' in galaxies at high-$z$. When compared to the estimates based on the cosmic star-formation history, current observations of galaxy populations at $z \sim 2.5$ are missing about half of the heavy element abundances expected \citep{bou07,gal08}. \textit{SPICA} would reveal the metal content of the dust-embedded star formation, which is missing in optically-based abundance measurements \citep{san10}. Still, a large fraction of metals are possibly locked in the hot intracluster gas \citep{fer05,dav07}, which is chemically enriched with the gas removed from galaxies by processes such as outflows --\,driven by star formation or AGN\,-- or ram-pressure stripping. \textit{Athena}+ will be able to measure the abundances and distribution of metals in clusters from the core to the boundary of the virial regions \citep{nan13}. Furthermore, the study of the dust composition in galaxies using IR spectroscopy with \textit{SPICA} could be complemented with the dust mineralogy obtained with \textit{Athena}+ from the analysis of X-ray absorption edges produced by chemical elements trapped into dust grains \citep[e.g.][]{lee05,hof16,pal16}. Thus, the combination of these two observatories would provide a accurate measurement of the metal budget in galaxies and the ICM from $z \sim 3$ till the present time.

Additionally, chemical evolution models could be constrained by including the content of the molecular and neutral gas with the help of ALMA and SKA, respectively, as mentioned in Section\,\ref{spicaevol}. The latter will be of particular interest to understand the matter cycle and the build-up and evolution of the metallicity relations found from high-$z$ galaxies to the Local Universe (see Section\,\ref{fmr}). On the other hand, the combined work of \textit{SPICA} and ALMA could follow the highly successful synergy between \textit{Spitzer} and \textit{Herschel} at low-$z$ to cover both sides of the dust thermal emission. These facilities would provide dust masses and temperatures for distant galaxies \citep[see][]{gru17}, as \textit{Herschel} did on nearby galaxies and a few of the brighter galaxies at high-$z$ \citep[e.g.][]{mag12}.

\section{SUMMARY}\label{sum}

One of the major challenges in astronomy for the next $\sim 20$ years is to understand the cycle of matter within galaxies and identify the physical mechanisms driving their evolution. This task can only be addressed from a multi-wavelength point of view, and thus will require the use of a sensitive IR spectroscopic observatory combined with the main observatories available in the next decades including \textit{JWST}, the ELTs, \textit{Athena}+, ALMA, NOEMA, SKA, but also the survey capabilities of, e.g. \textit{Euclid} and \textit{WFIRST}. The contribution of \textit{SPICA} would be essential to study the chemical evolution of galaxies, which is one of the most important manifestations of the baryon cycle activity in galaxies. The past history of gas accretion, star formation, nuclear activity, and gas outflows leaves an imprint on the composition and content of heavy elements and dust in the ISM of galaxies. Thus, metals and dust are relics through which one can reconstruct the evolutionary path that galaxies have followed from their formation to the present time.

The $2.5\, \rm{m}$ actively cooled mirror of \textit{SPICA} ($\sim 6\, \rm{K}$) would reach a unique sensitivity allowing the observation of the whole suite of lines and dust features in the rest-frame mid- and far-IR range, from the faintest regions in nearby galaxies to the knee of the luminosity function at the peak of star formation and black hole accretion activity ($1 < z < 3$). \textit{SPICA} would play a major role to unveil the dark side of galaxy evolution, since the bulk of star formation and black hole accretion can only be attained at long IR wavelengths \citep{spi17}. While \textit{JWST} will probe the external parts of high-$z$ galaxies through rest-frame optical/UV spectra, and ALMA will study the cold gas reservoir, the contribution of \textit{SPICA} would be essential to provide an insight into the warm gas in the dust-enshrouded regions of galaxies. No other facility with similar capabilities is currently planned. Therefore, rest-frame mid- to far-IR spectroscopy is essential to understand the physical processes at the engine of the matter cycle in galaxies.


Only a deep survey of spectroscopic observations with a sensitive IR observatory like \textit{SPICA} would reveal the chemical enrichment process for galaxies in the last $\sim 11\, \rm{Gyr}$. When combined with present and forthcoming first-line facilities in the decade of the 2020s, this spectroscopic survey would allow us to determine the evolutionary path followed by galaxies from the peak of star formation and SMBH accretion activity till the present time, and ultimately understand both the sequence of physical mechanisms involved in the matter cycle and how they drive the evolution of galaxies over cosmic time.

In this work we evaluate the feasibility of the main observational strategies foreseen to derive heavy element abundances and trace the dust composition in galaxies accessible to \textit{SPICA}, from high-$z$ to the Local Universe. IR diagnostics offer a huge advantage over optical/UV line ratios, since IR line intensities show a feeble response to both extinction by dust and gas temperature. Under the light of these tools, we also explore the potential contribution of an observatory with the characteristics of \textit{SPICA} to address some of the main questions in the field of galaxy evolution that will be unanswered in the late 2020s: how are galaxies assembled? How is the build-up of metals and dust? And how does the ISM composition evolve with redshift? The use of IR diagnostics applied to large spectroscopic samples would yield a unique view of the chemical evolution in galaxies, a key contribution to understand the picture provided by future studies based on \textit{JWST}, ALMA, and ELTs observations in the 2020s.

\begin{acknowledgements}
This paper is dedicated to the memory of Bruce Swinyard, who initiated the \textit{SPICA} project in Europe, but unfortunately died on 22 May 2015 at the age of 52. He was \textit{ISO}-LWS calibration scientist, \textit{Herschel}-SPIRE instrument scientist, first European PI of \textit{SPICA} and first design lead of SAFARI.

The SAFARI Consortium and the full \textit{SPICA} Team are acknowledged, without their work this mission project would not have been possible. J.A.F.O. acknowledges financial support from the Spanish Ministry of Economy and Competitiveness (MINECO) under grant number MEC-AYA2015-53753-P.
\end{acknowledgements}

\begin{appendix}

\section*{Affiliations}
\affil{$^1$Instituto de Astrof\'isica de Canarias (IAC), C/V\'ia L\'actea s/n, E--38205 La Laguna, Spain}%
\affil{$^2$Universidad de La Laguna (ULL), Dept. de Astrof\'isica, Avd. Astrof\'isico Fco. S\'anchez s/n, E--38206 La Laguna, Spain}%
\affil{$^3$Istituto di Astrofisica e Planetologia Spaziali (INAF--IAPS), Via Fosso del Cavaliere 100, I--00133 Roma, Italy}
\affil{$^4$IPAC, California Institute of Technology, Pasadena, CA 91125, USA}
\affil{$^5$Sterrenkundig Observatorium, Universiteit Gent, Krijgslaan 281 S9, 9000 Gent, Belgium}
\affil{$^6$Department of Physical Sciences, The Open University, MK7 6AA, Milton Keynes, United Kingdom}
\affil{$^7$Department of Astronomy and Joint Space Institute, University of Maryland, College Park, MD 20642 USA}
\affil{$^8$Laboratoire d'Astrophysique de Bordeaux, Univ. Bordeaux, CNRS, B18N, all\'ee Geoffroy Saint-Hilaire, 33615 Pessac, France}
\affil{$^9$Laboratoire AIM, CEA/IRFU/Service d'Astrophysique, Universit\'e Paris Diderot, Bat. 709, F--91191 Gif-sur-Yvette, France}
\affil{$^{10}$Department of Physics and Astronomy, University College London, Gower Street, London WC1E 6BT, UK}
\affil{$^{11}$Steward Observatory, University of Arizona, 933 North Cherry Avenue, Tucson, AZ 85721, USA}
\affil{$^{12}$Naval Research Laboratory, Remote Sensing Division, 4555 Overlook Avenue SW, Washington DC 20375, USA}
\affil{$^{13}$CNRS, IRAP, 9 Av. colonel Roche, BP 44346, 31028 Toulouse Cedex 4, France}
\affil{$^{14}$Universit\'e de Toulouse, UPS-OMP, IRAP, 31028 Toulouse Cedex 4, France}
\affil{$^{15}$Universidad de Alcal\'a, Dept. de F\'isica y Matem\'aticas, Campus Universitario, E-28871 Alcal\'a de Henares, Madrid, Spain}
\affil{$^{16}$Osservatorio Astronomico di Trieste, INAF, Via Tiepolo 11, I--34131 Trieste, Italy}
\affil{$^{17}$Osservatorio Astronomico di Bologna, INAF, via Ranzani 1, I--40127 Bologna, Italy}
\affil{$^{18}$National Astronomical Observatory of Japan, 2-21-1 Osawa, Mitaka, Tokyo 181-8588, Japan}
\affil{$^{19}$Graduate School of Science, Nagoya University, Furo-cho, Chikusa-ku, Nagoya 464-8602, Japan}
\affil{$^{20}$Astronomy Division, University of California, Los Angeles, CA 90095-1547, USA}
\affil{$^{21}$Institute of Space Astronautical Science, Japan Aerospace Exploration Agency, Sagamihara, Kanagawa 252-5210, Japan}
\affil{$^{22}$School of Physics and Astronomy, Cardiff University, Queen’s Buildings, The Parade, Cardiff CF24 3AA, United Kingdom}
\affil{$^{23}$Research Center for Space and Cosmic Evolution, Ehime University, Matsuyama 790-8577, Japan}
\affil{$^{24}$Centro de Astrobiolog\'ia (CSIC/INTA), Ctra. de Ajalvir km. 4, 28850 Torrej\'on de Ardoz, Madrid, Spain}
\affil{$^{25}$Department of Astronomy, Graduate School of Science, The University of Tokyo, 113-0033 Tokyo, Japan}
\affil{$^{26}$Department of Physics, University of Oxford, Oxford OX1 3RH, UK}
\affil{$^{27}$SRON Netherlands Institute for Space Research, Postbus 800, 9700, AV Groningen, The Netherlands}
\affil{$^{28}$Kapteyn Astronomical Institute, University of Groningen, Postbus 800, 9700 AV, Groningen, The Netherlands}
\affil{$^{29}$Osservatorio Astronomico di Roma, INAF, Via di Frascati 33, I--00078 Monte Porzio Catone, Italy}
\affil{$^{30}$Ritter Astrophysical Research Center, University of Toledo, 2825 West Bancroft Street, M. S. 113, Toledo, OH 43606, USA}
\affil{$^{31}$International Research Fellow of the Japan Society for the Promotion of Science (JSPS), Department of Astronomy, University of Tokyo, Bunkyo-ku, 113-0033 Tokyo, Japan}

\end{appendix}

\small
\bibliographystyle{pasa-mnras}
\bibliography{metaldust}

\end{document}